\documentstyle[11pt,paspconf,psfig]{article}

\begin{document}

\title{The large--scale distribution of galaxies in the Shapley Concentration}

\author{S. Bardelli}
\affil{Osservatorio Astronomico di Trieste, via Tiepolo 11, I--34131 Trieste, 
Italy}

\author{E. Zucca, G. Zamorani}
\affil{Osservatorio Astronomico di Bologna, via Zamboni 33, I--40126 Bologna, 
Italy}

\begin{abstract}
We present the results of a galaxy redshift survey in the central region of the 
Shapley Concentration. Our total sample contains $\sim 2000$ radial velocities 
of galaxies both in the clusters and in the inter-cluster field. We 
reconstruct the density profile of this supercluster, calculate its
overdensity and total mass. Moreover we detect a massive structure behind
the Shapley Concentration, at $\sim 30000$ km/s.
\end{abstract}


\keywords{Cosmology: observations - Cosmology: large-scale structure of the 
          Universe}

\section{Introduction}
The Shapley Concentration stands out as the richest system of Abell clusters
in the list of Zucca et al. (1993), at every density excess. 
In particular, at a density contrast of $\sim 2$, it has 25 members (at mean
velocity $\sim 14000$ km/s), while at the same density contrast the Great 
Attractor, which is the largest mass condensation within $80$ h$^{-1}$ Mpc,
has only 6 members and the Corona Borealis and Hercules superclusters are 
formed by 10 and 8 clusters, respectively.
While the cluster distribution in the Shapley Concentration has been
well studied, little is known about the distribution of the galaxies. 
Determining the properties of these galaxies is very important in order to 
assess the physical reality and extension of the structure and to determine 
if galaxies and clusters trace the matter distribution in the same way. 
 
\begin{figure}[t]
\centerline{{
\psfig{figure=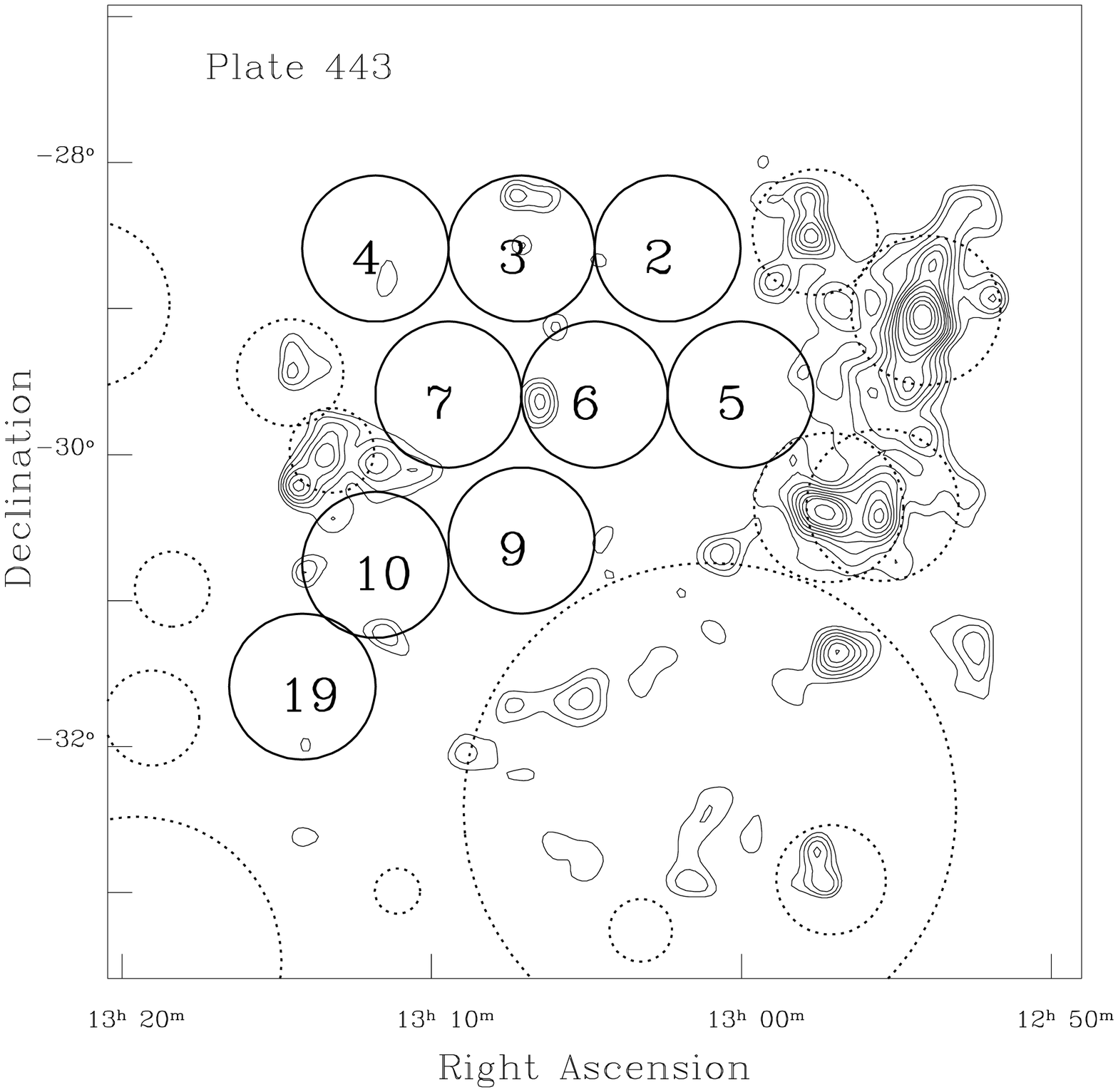,width=0.5\hsize} 
\psfig{figure=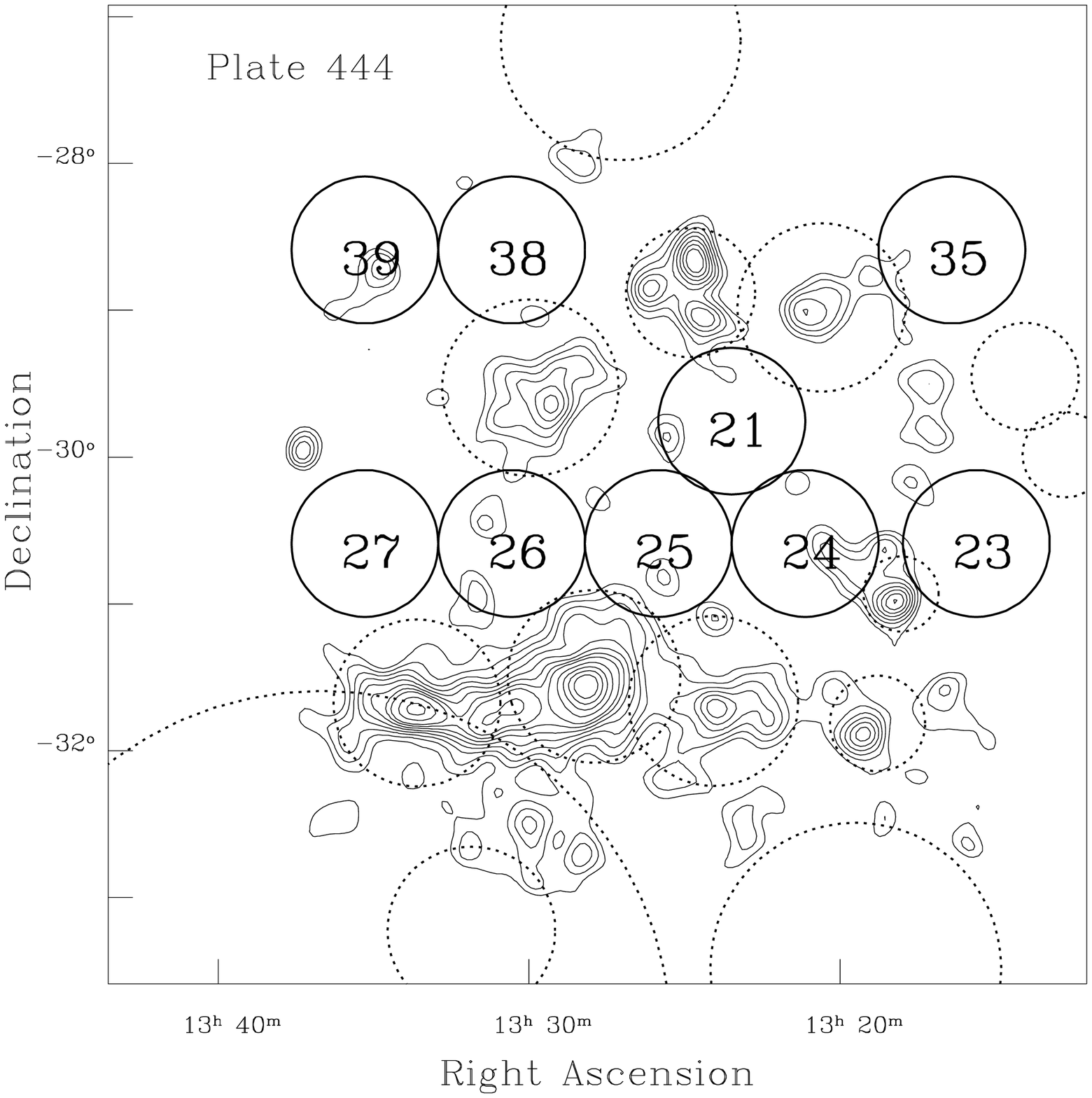,width=0.5\hsize} 
}}
\caption[]{Survey strategy for the observations of the inter-cluster galaxies 
in the UKSTJ plates 443 and 444. For the Abell clusters present in the plates, 
circles of one Abell radius have been drawn (dashed curves). Solid circles 
represent the MEFOS fields. 
}
\end{figure}

\begin{figure}
\centerline{\vbox{
\psfig{figure=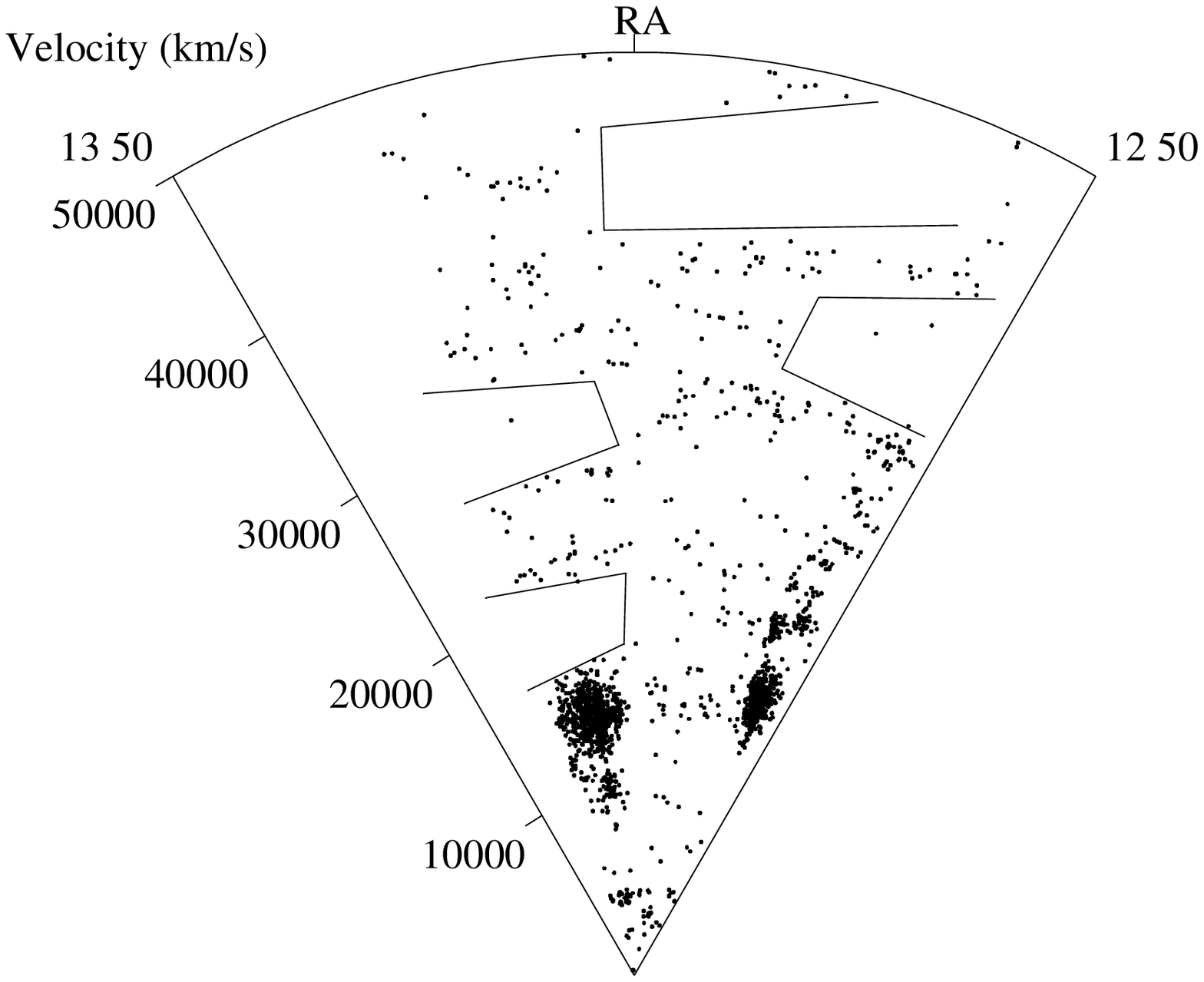,height=8.5cm}
}}
\centerline{\vbox{
\psfig{figure=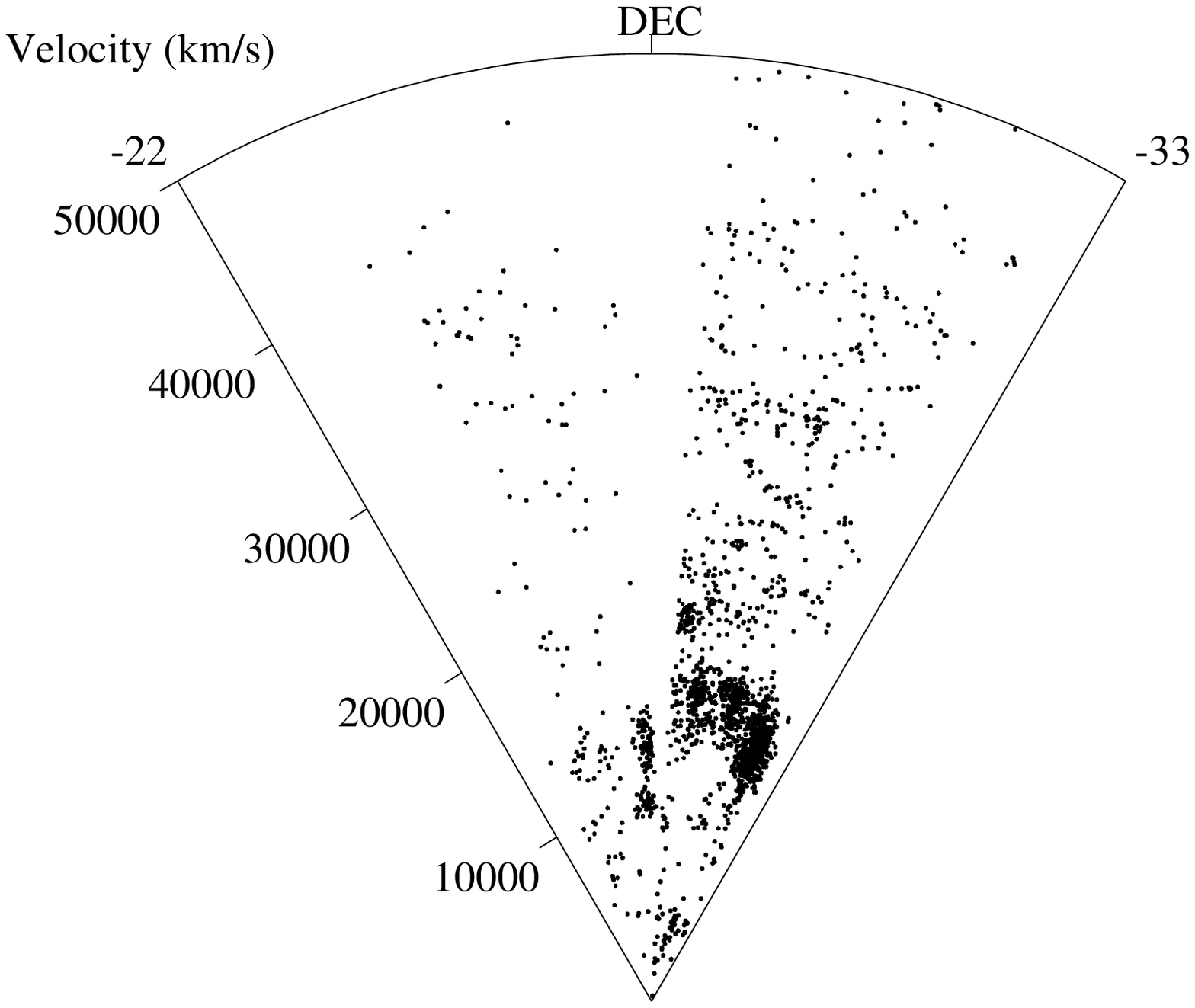,height=8.5cm}
}}
\caption[]{
Wedge diagrams of the galaxies in the velocity range $[0-50000]$ km/s.  
Upper panel: 1728 galaxies in the plates 443 and 444; note the clear presence 
of the A3558 (on the bottom left) and A3528 (on the bottom right) complexes 
and of four voids.
Lower panel: 1979 galaxies in the plates 443, 444 and 509; note that the  
plotted coordinate is the declination.
}
\end{figure}

\begin{figure}[t]
\centerline{\vbox{
\psfig{figure=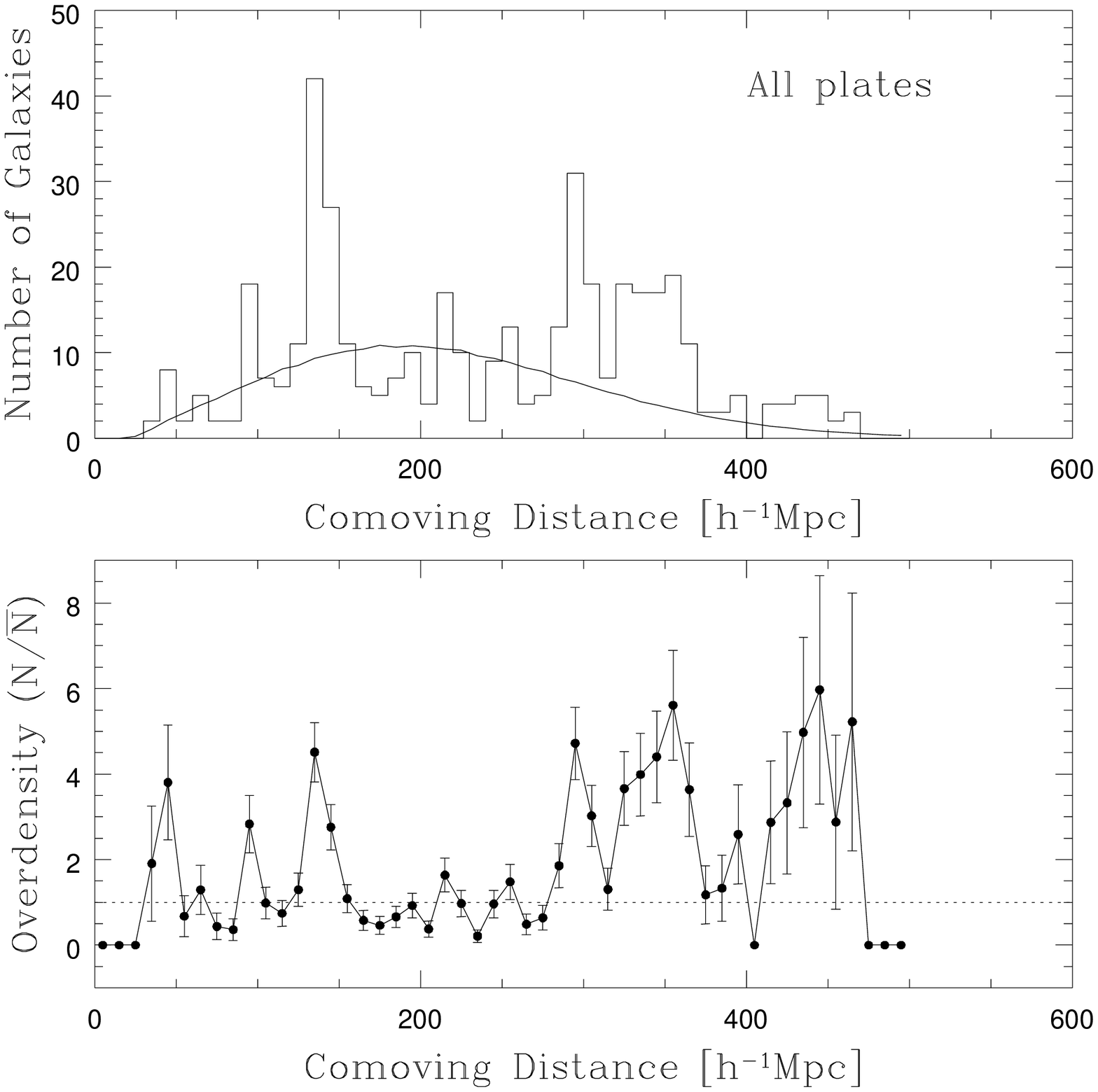,height=10.cm}
}}
\caption[]{Upper panel: Histogram of the velocities of galaxies between
clusters, with superimposed the expected distribution for a uniform sample. 
Lower panel: Galaxy density profile; the dashed line corresponds to an
overdensity $=1$. Galaxies in clusters are not included in these figures. 
}
\end{figure}

\section{The survey}

With the use of the OPTOPUS and MEFOS multifibre spectrographs at the 3.6m
ESO telescope we performed a redshift survey of galaxies in the two
cluster complexes (dominated by A3528 and A3558; Bardelli et al. 1994, 1998a,
1998b) at the center of the supercluster and of galaxies between clusters 
(the inter--cluster galaxies; Bardelli et al. 1999). 
The sampled area is contained in the UKSTJ plates 443, 444 and 509, 
corresponding to a scale of $\sim 25$ h$^{-1}$ Mpc at the distance of the 
Shapley Concentration, and our total sample contains 2057 velocities. 
In Fig.1 we show the observing strategy for the inter--cluster galaxies survey 
in plates 443 and 444, while in Fig.2 we report the wedge diagrams of the 
galaxy distribution. In the upper panel of Fig.2, it is clear the presence
of the two cluster complexes dominated by A3558 (on the left) and
by A3528 (on the right). These two structures appear to be connected by a 
bridge of galaxies, resembling the Coma-A1367 system in the central part 
of the Great Wall. The scale of this system is $\sim 23$ h$^{-1}$ Mpc and it is
similar to that of Coma-A1367 ($\sim 21$ h$^{-1}$ Mpc). Note also the presence
of two voids at $\sim 20000$ km/s and $\sim 30000$ km/s in the easternmost half 
of the wedge. Other two voids are visible in the westernmost part of the 
sample, delimiting two filamentary structures at $\sim 30000$ 
km/s and  $\sim 45000$ km/s, which connect each other at right ascension of 
$\sim 13^h$.

Knowing the incompleteness of the sample and using the luminosity function of 
field galaxies (from the ESP survey, Zucca et al. 1997), it is possible to 
reconstruct the density profile. In Fig.3 the histogram and the density 
profile of the 512 inter--cluster galaxies are shown. 
In the lower panel it appears clearly the presence of the large overdensity 
at $v=30000$ km/s. Note however that in this plot the Shapley Concentration
is not fully represented, lacking the large contribution of cluster galaxies.
The same analysis has been done for the
cluster complexes surveys and then the total density excess has been calculated.

The total overdensity in galaxies of this supercluster is $7.67\pm 1.91$
on a scale of $\sim 11$ h$^{-1}$ Mpc of radius, corresponding to a mass 
of $1.3\times 10^{16}$ h$^{-1}$ $M_{\odot}$ if light traces mass and $\Omega=1$. 
If the light
is more clustered than the mass and considering reasonable bias factors in the 
range $[1.4-2.5]$ (Hudson 1993), we find that the mass could lie in the range
$[2.73-4.62]\times 10^{15}$ h$^{-1}$ $M_{\odot}$.

If light traces mass and $\Omega=1$, the Shapley Concentration
 already stopped its expansion due to the Hubble flow and started to grow. 
Moreover, the central cluster 
complexes result to be in the collapse phase.
The predicted peculiar velocities inside the Shapley Concentration
are $\sim 1500$ km/s in the light--traces--mass case or in the range $[500-960]$
km/s in the case of biased galaxy formation.

For what concerns the structure at $\sim 30000$ km/s, its
estimated overdensity is $3.12$ on scales of $\sim 25$
h$^{-1}$ Mpc, corresponding to a mass of $5.96\times 10^{16}$ h$^{-1}$ 
$M_{\odot}$ if the light is an unbiased tracer of the underlying mass.



\begin{references}

\reference Bardelli S., Zucca E., Vettolani G., et al., 1994, \mnras, 267, 665 
 
\reference Bardelli S., Zucca E., Zamorani G., et al., 1998a, \mnras, 296, 599
 
\reference Bardelli S., Pisani A., Ramella M., et al., 1998b, \mnras, 300, 589  
 
\reference Bardelli S., Zucca E., Zamorani G., 1999, in preparation

\reference Hudson M.J., 1993, \mnras, 265, 43

\reference Zucca E., Zamorani G., Scaramella R., Vettolani G., 1993, 
           \apj, 407, 470

\reference Zucca E., et al., 1997, \aap, 326, 477


\end{references}
\end{document}